# Stabilization and modulation of the topological magnetic phase with a $Z_2$-vortex lattice in the Kitaev-Heisenberg honeycomb model: The key role of the third-nearest-neighbor interaction

Xiaoyan Yao* and Shuai Dong[†]

*School of Physics, Southeast University, Nanjing 211189, China*

[Abstract] The topologically nontrivial magnetic phase with a $Z_2$-vortex ($Z_2$V) lattice is investigated by simulation in the Kitaev-Heisenberg honeycomb model expanded by considering the second- and third-nearest-neighbor Heisenberg interactions ($J_{H2}$ and $J_{H3}$). On the parameter region of the $Z_2$V phase, a gradual modulation of vortex density is observed, together with a transition from single-$Z_2$V to triple-$Z_2$V state driven by the variation of frustration. Additionally, $Z_2$ vortices are arranged in different manners on the whole honeycomb structure for these two types of $Z_2$V states. Moreover, topologically equivalent states are revealed to exist in single-$Z_2$V dominant and triple-$Z_2$V dominant styles on different parameter points, which can be controlled to switch between each other without energy consumption. It is worth noting that $J_{H3}$ plays a key role in expanding the $Z_2$V phase, and also in stabilizing the single-$Z_2$V state.

* Corresponding author: yaoxiaoyan@seu.edu.cn

† Corresponding author: sdong@seu.edu.cn

I. INTRODUCTION

Vortices have long been a hot topic in many fields of physics research, and are believed to be associated with various intriguing phenomena, such as Abrikosov vortices in type-II superconductors [1, 2] and the vortex domains in multiferroic hexagonal manganites [3-7]. It is well known that the topological Z vortices characterized by the integral winding number Z could appear in a two-dimensional system where the order-parameter space is the two-dimensional rotation group SO(2) or the one-dimensional circle $S_1$ [SO(2)=$S_1$]. The Kosterlitz-Thouless transition in a XY ferromagnet is driven by the dissociation of Z vortices [8, 9]. If the ferromagnet is considered to be Heisenberg type, the Z vortex will be topologically trivial because the order-parameter space is the two-dimensional sphere $S_2$. Since there is no topological point defect, no topological phase transition takes place. However, the frustration could bring another type of topological defect into Heisenberg spin systems. When the order-parameter space is the three-dimensional rotation group SO(3) or the projective space $P_3$ [SO(3)=$P_3$] as shown in the frustrated triangular Heisenberg antiferromagnet, a $Z_2$ vortex ($Z_2$V) characterized by the two-valued topological quantum number emerges. Here, the ground state is the 120º spin structure with 120º between every two nearest-neighbor spins on the triangular lattice. The order parameter is not one spin, but a rigid rotator consisting of three spins on an elementary triangle [10]. These rotators circulate around a core making a topologically stable vortex. Two $Z_2$ vortices ($Z_2$Vs) can annihilate each other. Whether they circulate in clockwise or counterclockwise fashion makes no difference topologically. The binding-unbinding of these thermally excited $Z_2$Vs drives the thermodynamic phase transition at a finite temperature without conventional long-range order in these systems [10-13].

Since the unconventional bond-dependent Kitaev coupling was believed to be substantial in the hexagonal iridates $A_2$IrO$_3$ ($A$ = Na, Li), the Kitaev-Heisenberg (KH) model attracted considerable attention, and had been proposed to capture the essential magnetic interactions in these iridates [14-22]. Recently, $\alpha$-RuCl$_3$ was also suggested as a candidate Kitaev material in a 4$d$ analog of honeycomb iridates [23-30]. The unconventional Kitaev interaction breaks the spin rotational symmetry and provides an avenue for a new kind of frustration, namely, the Kitaev frustration. As the geometrical frustration is introduced, the region with puzzled magnetic states was observed on the phase diagram [31, 32]. Recently, the topological $Z_2$Vs, not thermally excited, were observed in the triangular KH model. The Kitaev term in the exchange Hamiltonian condenses these vortices into a $Z_2$V crystal at zero temperature [33, 34]. In addition, our previous report explored a wide parameter range of the KH honeycomb model with the next-nearest-neighbor Heisenberg interaction considered, and found an inhomogeneous state exhibiting a topological triple-vortex lattice. But this triple-$Z_2$V state only exists on the phase boundaries [35]. Its existing region of the ground state is too narrow to be well located, and thus it will be too hard to be observed in experiments.

The present paper focuses on the topological $Z_2$V phase in the KH honeycomb model expanded by considering further neighboring interactions. The existing range of the $Z_2$V ground state is carefully confirmed by detailed analyses, and it is revealed to be effectively expanded by the third-nearest-neighbor Heisenberg interaction with appropriate magnitude. In the $Z_2$V phase area, a nearly continuous modulation of vortex density is observed, resulting from a continuous modulation on spin configurations. Furthermore, a transition from single-$Z_2$V to triple-$Z_2$V state is revealed to be induced by the variation of mixed frustration, and the third-nearest-neighbor Heisenberg interaction plays a key role in stabilizing the single-$Z_2$V state. The topologically equivalent states could exist in single-$Z_2$V dominant and triple-$Z_2$V dominant fashions on different parameter points, which can be controlled to switch between each other without energy consumption. Moreover, the simulation shows that the $Z_2$Vs are arranged in different styles on the honeycomb structure for the single-$Z_2$V and triple-$Z_2$V states.

II. METHODS

Considering the expanded KH model where $J_K$ is the Kitaev interaction coupling different spin components ($S^x$, $S^y$, and $S^z$) on the nearest-neighbor bonds along the three lattice directions labeled by $\gamma=xx$, $yy$ and $zz$ as presented in Fig. 1(a). $J_H$, $J_{H2}$, and $J_{H3}$ represent the isotropic Heisenberg couplings between spins on the nearest-neighbor ($<i, j>$), the second-nearest-neighbor ($<<i, k>>$), and the third-nearest-neighbor ($<<<i, l>>>$) sites. The Hamiltonian takes the form of

$$H = J_K \sum_{<i,j>_\gamma} S_i^\gamma \cdot S_j^\gamma + J_H \sum_{<i,j>} S_i \cdot S_j + J_{H2} \sum_{<<i,k>>} S_i \cdot S_k + J_{H3} \sum_{<<<i,l>>>} S_i \cdot S_l . \quad (1)$$

$\sqrt{J_H^2 + J_K^2} = 1$ is fixed as the energy scale. We consider the region of antiferromagnetic (AFM) $J_H>0$ and ferromagnetic $J_K<0$ as originally proposed for the KH model relevant to iridates [14, 15]. Then the ratio of $J_H$ to $J_K$ is considered by parametrizing $J_H=\cos\varphi$ and $J_K=\sin\varphi$ with $\varphi$: $1.5\pi\sim2\pi$. Monte Carlo (MC) simulation of the Metropolis algorithm combined with the over-relaxation method is performed on the honeycomb lattice of $N=9216$ sites with periodic boundary conditions assumed [36, 37]. On every parameter point, the system is first evolved from a relatively high temperature ($T$) near 1 to a very low $T=0.0000001$ gradually, and then the energy is further minimized by 100000 MC steps restricted at $T=0$ (namely, only the proposed update with the energy variation not higher than zero is accepted) to approach the limit of zero temperature. The final result is obtained by comparing independent data sets evolving from different initial states.

The states obtained from MC simulation can be preliminarily distinguished by the spin correlations as calculated by

$$\begin{aligned} C_n &= \langle S_i \cdot S_{i+1} \rangle_n \\ C_{nn} &= \langle S_i \cdot S_{i+2} \rangle_{nn} \\ C_{nnn} &= \langle S_i \cdot S_{i+3} \rangle_{nnn} \\ C_K &= \langle S_i^x \cdot S_{i+1}^x \rangle_{xx} + \langle S_i^y \cdot S_{i+1}^y \rangle_{yy} + \langle S_i^z \cdot S_{i+1}^z \rangle_{zz} \end{aligned} , \quad (2)$$

where $C_n$, $C_{nn}$, and $C_{nnn}$ are spin correlations on the nearest-neighbor, the second-nearest-neighbor, and the third-nearest-neighbor spin pairs. $C_K$ is the spin correlation of the nearest-neighbor Kitaev bonds. Five well-known commensurate ordered phases —ferromagnetic (F), Neel (N), stripy (S), zigzag (Z) and double 120° (D) phases—are identified by comparing to the standard values of correlations listed in Table I with less than 15% deviation. Here D is a phase where both triangular sublattices show coplanar 120° spin structures. These phases are further confirmed by the spin structure factors [38, 39], which can be calculated on one triangular sublattice (TSL) of the honeycomb structure as follows:

$$S^{\gamma}(\boldsymbol{k}) = \sum_{i,j} e^{i\boldsymbol{k}\cdot(\boldsymbol{r}_j - \boldsymbol{r}_i)} \langle S_i^{\gamma} \cdot S_j^{\gamma} \rangle_{TSL} \tag{3}$$

$S^{\gamma}(\boldsymbol{k})$ is evaluated for three spin components ($\gamma=x$, $y$, and $z$), respectively, to show the detailed spin structure induced by the anisotropic Kitaev interaction.

To distinguish the phase with $Z_2$Vs from the unidentified region, further analysis is required. First, since $Z_2$V is a topological defect of the 120º spin structure, the local 120º spin pattern with slight distortion is required. To measure it, the chirality vector ($\boldsymbol{\kappa}$) is calculated on the spin configuration of one TSL in the form of

$$\boldsymbol{\kappa}(r) = \frac{2}{3\sqrt{3}}(S_1 \times S_2 + S_2 \times S_3 + S_3 \times S_1), \tag{4}$$

where the subscripts 1, 2, and 3 number the corner sites clockwise around each elementary triangle [Fig. 1(a)]. The orientation of $\boldsymbol{\kappa}$ is perpendicular to the plane on which the local 120º spin pattern lies approximately. The magnitude of $\boldsymbol{\kappa}$ ($|\kappa|$) gives a measure of the rigidity of the 120° pattern, and it is normalized to unity for a perfect one. Here the averaged magnitude of $\boldsymbol{\kappa}$ higher than 0.74 is set for the $Z_2$V phase to ensure the 120° spin structure kept locally. Second, the typical feature of spin structure factor for the $Z_2$V phase, where 12 main peaks for three spin components are separately located along the hexagon's edges, is checked [35]. Third, the vorticity ($v$) is calculated on $\boldsymbol{\kappa}$ configuration in the same manner as Ref. [10] by going around the right- and left-pointing triangular loops as illustrated in Fig. 1(a). $v$ is rescaled to be zero for no rotation and 1 for a rotation by $2\pi$. The sum of $v$ (VS) is counted, and its nonzero value finally confirms the $Z_2$V phase. Following these steps, $Z_2$V can be well located and measured. The simulation confirms that $Z_2$V is actually the rotation of the local 120º pattern. As shown in Fig. 1(b), the calculated $\boldsymbol{\kappa}$ vectors form a typical vortex with the smallest $|\kappa|$ as the vortex core. Correspondingly, the calculated $v$ map confirms a $Z_2$V in the center as illustrated in Fig. 1(c). If two $Z_2$V cores are counted in one loop, $v=0$ is obtained, implying the annihilation of these two $Z_2$Vs.

To explore ground states, the averaged energies per site for the commensurate ordered F, N, S, Z, and D phases are estimated as follows:

$$\begin{aligned} E_F &= (3J_H + 6J_{H2} + 3J_{H3} + J_K)/2 \\ E_N &= (-3J_H + 6J_{H2} - 3J_{H3} - J_K)/2 \\ E_S &= (-J_H - 2J_{H2} + 3J_{H3} - |J_K|)/2 \\ E_Z &= (J_H - 2J_{H2} - 3J_{H3} - |J_K|)/2 \\ E_D &= (-3J_{H2} - |J_K|)/2 \end{aligned} \tag{5}$$

Every state obtained from MC simulation is checked by comparing its energy ($E_{MC}$) to the above energies of commensurate phases to confirm the ground state. The energy of the metastable state is required to be not higher than the ground state by 0.005.

III. RESULTS AND DISCUSSION

The phase diagrams obtained by simulation indicate that $J_{H2}$ and $J_{H3}$ play important roles in generating $Z_2V$ phase. Figures 2(a)-2(d) plot the phase diagrams in the space of $\varphi$ and $J_{H2}$ with different $J_{H3}$. Without $J_{H2}$ and $J_{H3}$, only S and N phases can be identified on the bottom line in Fig. 2(a), and the S phase is induced by Kitaev frustration. When the geometrical frustration is switched on by AFM $J_{H2}$, the D phase, which is the base to generate $Z_2$Vs, appears and spreads with $J_{H2}$ increasing. The interaction between Kitaev frustration and geometrical frustration induces the $Z_2V$ state appearing along the boundary between D and S phases, but its region as a ground state is too narrow as mentioned in Ref. [35] without $J_{H3}$. According to $C_{nnn}$ listed in Table I, the AFM $J_{H3}$ enhances Z and N phases, while suppressing S and F phases. When $J_{H3}$ is switched on, the Z phase spreads into this parameter region [Figs. 2(b-d)], and it is confirmed as the ground state for $Na_2IrO_3$ and $\alpha$-$RuCl_3$ [26, 27, 40-42]. At the same time, the obvious shrink of the S phase opens an interval between D and S phases, in which the $Z_2V$ phase expands greatly, as shown in Fig. 2(b) with $J_{H3}$=0.1. As $J_{H3}$ is increased to 0.15, the S phase nearly disappears and the $Z_2V$ phase is further enhanced [Fig. 2(c)]. However, as $J_{H3}$ is raised higher, the region of the $Z_2V$ phase is suppressed and moved to higher $\varphi$ and $J_{H2}$ with the expansion of N and Z phases [Fig. 2(d)]. On the other hand, Figs. 2(e)-2(h) plot the phase diagrams in the space of $\varphi$ and $J_{H3}$ with different $J_{H2}$, where AFM $J_{H2}$ enhances the D phase but suppresses the N phase according to $C_{nn}$ listed in Table I. As $J_{H2}$ increases, firstly the $Z_2V$ phase spreads to some extent with the N phase shrinking, and then it is squeezed by too strong a D phase. In summary, the optimal parameter range of the $Z_2V$ phase is about 0.3-0.5 for $J_{H2}$ and 0.1-0.2 for $J_{H3}$. Besides $J_{H2}$, which brings geometrical frustration to produce the D phase, $J_{H3}$ plays a key role in stabilizing the $Z_2V$ phase on a larger parameter region. In addition, the metastable $Z_2V$ state always appears in the ground-state region of the D phase to the left of the $Z_2V$ ground state. As the topological point defects, $Z_2$Vs cannot be removed by small modifications, which makes them rather stable, existing in ground states and metastable states on the phase diagram of zero temperature.

In the whole region of the $Z_2V$ phase in the phase diagrams of Fig. 2, all $Z_2$Vs are observed in the form of a $Z_2V$ crystal. The vortex density increases with $\varphi$ rising but decreases with $J_{H2}$ or $J_{H3}$ rising, presenting a nearly continuous modulation. Figure 3 illustrates two typical cases with VS=1152 and 72, namely, a dense and a sparse $Z_2V$ crystals. As shown in Fig. 3(a), VS=1152 is the densest $Z_2V$ lattice that can be shown here, where $Z_2$Vs are closely packed in a Kagome lattice as presented in the $v(r)$ map. Higher VS exists, but it is too dense to be exactly located and exhibited by the present method. In the sparse case of VS=72 [Fig. 3(d)], $Z_2$Vs are separated far away from each other, forming a triangular lattice. In addition, $S^\gamma(\boldsymbol{k})$ shows the typical peaks as plotted in Figs. 3(b) and 3(e). As VS decreases, the main peaks of $S^\gamma(\boldsymbol{k})$ slide along the hexagon's edges from M to K points (M is the midpoint of the edge and K is the corner of the first Brillouin zone). The vortex comes from the superlattice of $\kappa(r)$, which can be better characterized by a Fourier transform, that is, the structure factor of $\kappa(r)$ [$S\kappa^\gamma(\boldsymbol{k})$] can be calculated by

$$S\kappa^\gamma(\boldsymbol{k}) = \sum_{i,j} e^{i\boldsymbol{k}\cdot(\boldsymbol{r}_j-\boldsymbol{r}_i)} \left\langle \kappa_i^\gamma \cdot \kappa_j^\gamma \right\rangle_{\text{TSL}}, \tag{6}$$

where $S\kappa^\gamma(\boldsymbol{k})$ is also calculated for three components ($\gamma$=$x$, $y$, and $z$), respectively. The main peaks of $S\kappa^\gamma(\boldsymbol{k})$ appear on the midperpendicular lines to the hexagon's edges, moving toward the center with decreasing VS, as shown in Figs. 3(c) and 3(f). If the wave vector of the $S^\gamma(\boldsymbol{k})$ main peak is denoted by $\boldsymbol{k}_p$ and that of the $S\kappa^\gamma(\boldsymbol{k})$ main peak is denoted by $\boldsymbol{kk}_p$, their averaged magnitudes $|k_p|$ and $|kk_p|$ present a good correspondence to VS as plotted in Fig. 4(a). With VS rising, $|k_p|$ decreases while $|kk_p|$ increases gradually, showing that the continuous variation of vortex density results from the continuous modulation on $\kappa(r)$ and spin configuration.

It is interesting that the modulation of the $Z_2V$ lattice is actually not a simple drifting apart. When VS is low, $Z_2Vs$ exist in the single style as shown in Fig. 3(d). As VS increases, the aggregation of $Z_2Vs$ prefers trimerization, and thus the triple-$Z_2V$ state with three $Z_2Vs$ combined together is observed as reported in the case without $J_{H3}$ [35]. To detect this detail, the ratio of single vortices to all vortices (RS) is estimated. In the case of $J_{H3}=0$ [Fig. 2(a)], only the $Z_2V$ states with RS near zero are displayed. When $J_{H3}=0.2$ [Fig. 2(d)], there are only the $Z_2V$ states with RS=1. As $J_{H3}=0.1$ and 0.15 [Figs. 2(b) and 2(c)], RS presents a more abrupt transition than the continuous variation of VS, which is also seen in Figs. 2(e)-2(h). The jump of RS means a transition from the single- to triple-$Z_2V$ state, and it happens at about $J_{H3}=0.1$-0.15. It is worth noting that the triple-$Z_2V$ state always appears beside the S phase. If the S phase disappears completely, the triple-$Z_2V$ state also vanishes as shown in Fig. 2(d). As plotted in Figs. 4(c)-4(f), from the single-$Z_2V$ state near the D phase to the triple-$Z_2V$ state close to the S phase on the $Z_2V$ area, $|C_n|$ and $|C_{nnn}|$ (correlations between two TSLs) increase, while $|C_{nn}|$ (correlation within one TSL) and $|\kappa|$ decrease. It is indicated that the single-$Z_2V$ state with stronger intra-TSL correlation and weaker inter-TSL ones mostly results from the mixed frustration within one TSL, while the triple-$Z_2V$ state is induced by the combined frustration from both intra-TSL and inter-TSL origins on the whole honeycomb structure. Here, $J_{H3}$ plays a crucial role in suppressing the S phase. When the range of $Z_2V$ phase broadens, $J_{H3}$ mediates the inter-TSL geometrical frustration, and thus stabilizes the single-$Z_2V$ state with a higher $|\kappa|$. ($C_{nnn}$ is positive but $C_n$ and $C_{nn}$ are negative in the $Z_2V$ phase.)

It is noteworthy that the variation of RS from zero to one does not depend on VS monotonously. Figure 4(b) plots RS dependence on VS of all the $Z_2V$ ground states. When VS is less than 420, RS equals 1. When VS is higher than 1152, RS approaches zero. But in the middle region, states with different RS exist at the same VS. The typical example is illustrated in Fig. 5. The first and second columns, respectively, plot the $Z_2V$ ground states with RS=1 and 0.2222 for the same VS=648. The third and fourth columns, respectively, plot a pure single-$Z_2V$ ground state with RS=1 and a pure triple-$Z_2V$ metastable state with RS=0 at the same VS=288. The single- and triple-$Z_2V$ features can be directly seen on the $v(r)$ map in Fig. 5. [The single-$Z_2Vs$ in Fig. 5(g) are separated equidistantly [43]]. In the case of RS=1, $S^\gamma(\boldsymbol{k})$ in Figs. 5(b) and 5(h) presents the typical peaks similar to those observed in the triangular KH model [33], implying a modulation of the triangular superlattice originated from the approximatively decoupled TSLs. Correspondingly $S\kappa^\gamma(\boldsymbol{k})$ of Figs. 5(c) and 5(i) shows the simple peaks similar to Fig. 3(f). In the case of RS=0.2222 and 0, $S^\gamma(\boldsymbol{k})$ and $S\kappa^\gamma(\boldsymbol{k})$ show the main peaks the same as the state of RS=1. However, besides main peaks, there are extra small peaks emerging on the hexagon's edges for $S^\gamma(\boldsymbol{k})$ and along some special lines for $S\kappa^\gamma(\boldsymbol{k})$ [Figs. 5(e), 5(f), 5(k), and 5(l)], which result from the extra hexangular character attached on the triangular superlattice in the honeycomb structure. This kind of triple-$Z_2V$ state is only observed in the KH honeycomb model. Since $Z_2Vs$ are topologically stable, extra energy is required to produce or remove $Z_2Vs$, namely, a potential barrier exists between the $Z_2V$ states with different VS. But for the $Z_2V$ states with the same VS, there is no potential barrier between them, namely, one $Z_2$V state can reach another one with the same VS without energy consumption by just changing parameters at $T$=0. These topologically equivalent $Z_2$V states show triple-$Z_2$V dominant and single-$Z_2$V dominant features on different parameter points due to different frustration conditions. There may be an effective force between $Z_2$Vs, which is decided by the coupling parameters, but its distance dependence is too complex to be written into a simple expression due to multiple frustrations.

Since the $Z_2$Vs are calculated on one TSL as mentioned above, an interesting question remains, that is, how the $Z_2$Vs on two TSLs are arranged in the whole honeycomb structure. The simulation reveals that they are arranged in different styles for the single-$Z_2$V and triple-$Z_2$V states. In the case of the single-$Z_2$V state, the single-$Z_2$Vs emerge as typical $\kappa$ vortices on (1,1,1), (1,-1,-1), (-1,1,-1), and (-1,-1,1) planes (marked as 1, 2, 3, and 4), forming a triangular lattice on each TSL as illustrated in Fig. 6(a). If these single-$Z_2$Vs on two TSLs are plotted into one picture as given in Fig. 6(b), these two single-$Z_2$V triangular lattices interpenetrate into each other to construct a honeycomb lattice in the manner that each single-$Z_2$V has three neighbors on different planes. On the other hand, in the case of the triple-$Z_2$V state, the $\kappa(r)$ map on one TSL shows hexagonal domain structure, where $\kappa$ domains along [1,1,1] (or [-1,-1,-1]), [1,-1,-1] (or [-1,1,1]), [-1,1,-1] (or [1,-1,1]), and [-1,-1,1] (or [1,1,-1]) are marked as 1, 2, 3, and 4. The $Z_2$V triples appearing on the intersection of six domains of three types are also marked as 1, 2, 3, and 4 in the form of Fig. 6(e), i.e., they are named after the type of $\kappa$ domain that does not appear around this $Z_2$V triple. These $Z_2$V triples construct a triangular lattice on each TSL. If $\kappa(r)$ maps on two TSLs are plotted into one picture as illustrated in Fig. 6(f), the $Z_2$V triples of the same type from two TSLs nearly overlap with each other, and moreover the $\kappa$ domains of the same type and nearly opposite orientations also overlap with each other. Although $\kappa(r)$ and $v(r)$ maps are calculated on one TSL, they demonstrate the nontrivial modulation actually existing on the whole honeycomb structure. As plotted in Figs. 6(c) and 6(g), the site-dependent local energy $E_i(r)$ of the honeycomb structure just displays the pattern corresponding to the $\kappa(r)$ map including two TSLs. All these nontrivial properties finally come from the modulation of spin configuration on the whole honeycomb structure, which presents a very complex superlattice pattern as plotted in Figs. 6(d) and 6(h).

IV. CONCLUSION

In summary, the $Z_2$V phase is well located and systematically investigated by simulation in the KH honeycomb model expanded by considering $J_{H2}$ and $J_{H3}$. In the region of the $Z_2$V phase, a gradual modulation of vortex density is observed, together with a transition from single-$Z_2$V to triple-$Z_2$V state driven by the variation of frustration condition. The triple-$Z_2$V state is induced by the frustration from both intra-TSL and inter-TSL origins on the whole honeycomb structure, while the single-$Z_2$V state mostly results from the frustration within one TSL. Topologically equivalent states with the same VS, showing the single-$Z_2$V dominant and the triple-$Z_2$V dominant features on different parameter points, can be controlled by the frustration condition to switch between each other without energy consumption. For the whole honeycomb structure, the $Z_2$Vs are arranged in different styles, namely, interpenetrating for the single-$Z_2$V state and overlapping for the triple-$Z_2$V state. It is worth noting that $J_{H3}$ is a crucial factor to the $Z_2$V phase in this system. It plays a key role in expanding and stabilizing the $Z_2V$ phase. Moreover it mediates the inter-TSL geometrical frustration and thus modulates the single-$Z_2V$ state.

**Acknowledgments**

This work is supported by research grants from the National Natural Science Foundation of China (Grant Nos. 11674055). We thank Prof. Maria Daghofer for helpful discussions.

Figure Legends:

FIG. 1. (a) A sketch of the honeycomb lattice, which is composed of two TSLs as represented by white and gray dots, respectively. The solid, dashed, and dotted black thick lines indicate three kinds of spin-dependent nearest-neighbor bonds, where $xx$, $yy$ and $zz$ involve $S^x$, $S^y$, and $S^z$, respectively. The vector chirality ($\kappa$) is calculated on the shading downward-pointing elementary triangles of one TSL. The vorticity ($v$) is evaluated on right- and left-pointing triangular loops connecting one-third downward-pointing triangles with darker shading. (b) Partial $\kappa(r)$ map, where the arrows present the projections of $\kappa$ onto the (111) plane, and the color refers to the value of $|\kappa|$. (c) Partial $v(r)$ map with one $Z_2V$ calculated on the $\kappa(r)$ configuration of (b), where the yellow dot represents $v=1$ and the black one denotes $v=0$.

FIG. 2. The phase diagram in the space of $\varphi$ and $J_{H2}$ for (a) $J_{H3}=0$, (b) $J_{H3}=0.1$, (c) $J_{H3}=0.15$, and (d) $J_{H3}=0.2$, and in the space of $\varphi$ and $J_{H3}$ for (e) $J_{H2}=0.2$, (f) $J_{H2}=0.3$, (g) $J_{H2}=0.4$, and (h) $J_{H2}=0.5$. The solid squares filled with different colors show the ground states identified as the commensurate ordered phases, and those empty squares represent the unidentified regions. The large solid circles show $Z_2V$ ground states, and the solid small circles on colorful squares show $Z_2V$ metastable states, where the shading refers to the value of VS. The shading of the small filled triangles on the circles refers to the value of RS.

FIG. 3. (a) Partial map of $v(r)$, (b) $S^\gamma(\mathbf{k})$, and (c) $S\kappa^\gamma(\mathbf{k})$ in the ground state of VS=1152 obtained with $J_{H2}=0.3$, $J_{H3}=0.05$, and $\varphi=1.65625\pi$. (d) Partial map of $v(r)$, (e) $S^\gamma(\mathbf{k})$, and (f) $S\kappa^\gamma(\mathbf{k})$ in the ground state of VS=72 obtained at $J_{H2}=0.3$, $J_{H3}=0.3$, and $\varphi=1.84375\pi$. For the $v(r)$ map, the yellow dot represents $v=1$ and the black one denotes $v=0$. For $S^\gamma(\mathbf{k})$ and $S\kappa^\gamma(\mathbf{k})$, the solid, dashed and dotted lines with arrows denote the signals from $x$, $y$ and $z$ components, respectively.

FIG. 4. (a) $|k_p|$ and $|kk_p|$ as functions of VS with error bars. (b) The dependence of RS on VS. (c,d) The $Z_2V$ phase area of Fig. 2(b) with $J_{H3}=0.1$. (e,f) The $Z_2V$ phase area of Fig. 2(g) with $J_{H2}=0.4$. (c,e) The shading of the left-half dot refers to the value of $|C_n|$ from 0.280(white) to 0.872(dark), and that of the right half refers to the value of $|C_{nnn}|$ from 0.350(white) to 2.350(dark). (d,f) The shading of the left-half dot refers to the value of $|C_{nn}|$ from 2.188(white) to 2.692(dark), and that of the right half refers to the averaged value of $|\kappa|$ from 0.759(white) to 0.905(dark).

FIG. 5. Partial $v(r)$ map, $S^\gamma(\mathbf{k})$, and $S\kappa^\gamma(\mathbf{k})$ are displayed in the first, second, and third rows. The first column gives the ground state with VS=648 and RS=1 obtained at $J_{H2}=0.45$, $J_{H3}=0.15$ and $\varphi=1.84375\pi$. The second column shows the ground state with VS=648 and RS=0.22222 obtained at $J_{H2}=0.5$, $J_{H3}=0.0$ and $\varphi=1.65625\pi$. The third column gives the ground state with VS=288 and RS=1 obtained at $J_{H2}=0.2$, $J_{H3}=0.2$ and $\varphi=1.75\pi$. The fourth column presents the metastable state with VS=288 and RS=0 obtained at $J_{H2}=0.45$, $J_{H3}=0.0$ and $\varphi=1.625\pi$. For the $v(r)$ map, the yellow dot represents $v=1$ and the black one denotes $v=0$. For $S^\gamma(\mathbf{k})$ and $S\kappa^\gamma(\mathbf{k})$, the solid, dashed, and dotted lines with arrows denote the signals from $x$,

*y*, and *z* components, respectively.

FIG. 6. The first row displays the single-$Z_2$V ground state with VS=72 and RS=1 obtained at $J_{H2}$=0.3, $J_{H3}$=0.3, and $\varphi$=1.84375$\pi$. The second row shows the triple-$Z_2$V metastable state with VS=288 and RS=0 obtained at $J_{H2}$=0.45, $J_{H3}$=0.0, and $\varphi$=1.625$\pi$. The first column presents the $\kappa(r)$ map on one TSL. The second column displays the $\kappa(r)$ map with all $\kappa$ vectors from two TSLs. (See Supplemental Material for details [43].) The arrows present the projections of $\kappa$ onto the (111) plane, and the color refers to the value of |$\kappa$|. The single-$Z_2$Vs are marked as 1, 2, 3 and 4 written on solid circles in (a). In (b), besides the $Z_2$Vs from (a), the single-$Z_2$Vs from another TSL are also marked in the same manner as 1, 2, 3, and 4 written on solid rectangles. For (e) and (f), $\kappa$ domains are marked as 1, 2, 3, and 4. The $Z_2$V triples are marked as 1, 2, 3, and 4 written on solid circles. The third column exhibits the $E_i(r)$ map of the honeycomb structure, and the shading refers to the value of $E_i$. The fourth column plots the spin configuration of the honeycomb structure. The vectors show the projections of spins onto the (111) plane with the shading referring to the component of spin perpendicular to the (111) plane. For visibility, only part of the lattice is plotted.

Table I. The standard values of the correlation functions $C_n$, $C_{nn}$, $C_{nnn}$, and $C_K$ for F, N, S, Z, and D phases.

|         | F | N  | S  | Z  | D  |
|---------|---|----|----|----|----|
| $C_n$   | 3 | -3 | -1 | 1  | 0  |
| $C_{nn}$| 6 | 6  | -2 | -2 | -3 |
| $C_{nnn}$| 3 | -3 | 3  | -3 | 0  |
| $C_K$   | 1 | -1 | 1  | 1  | 1  |

Fig. 1

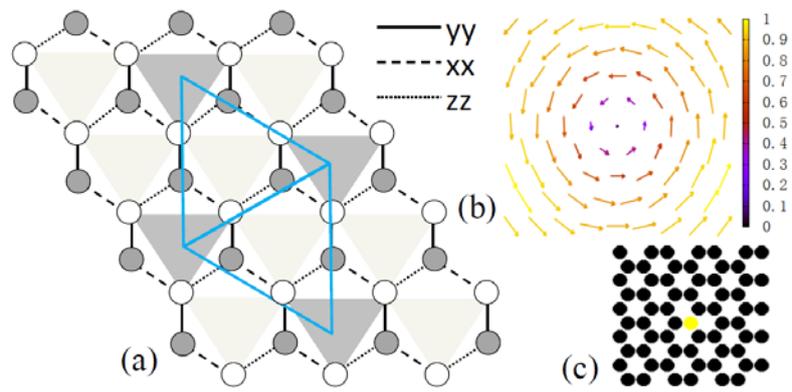

Fig. 2

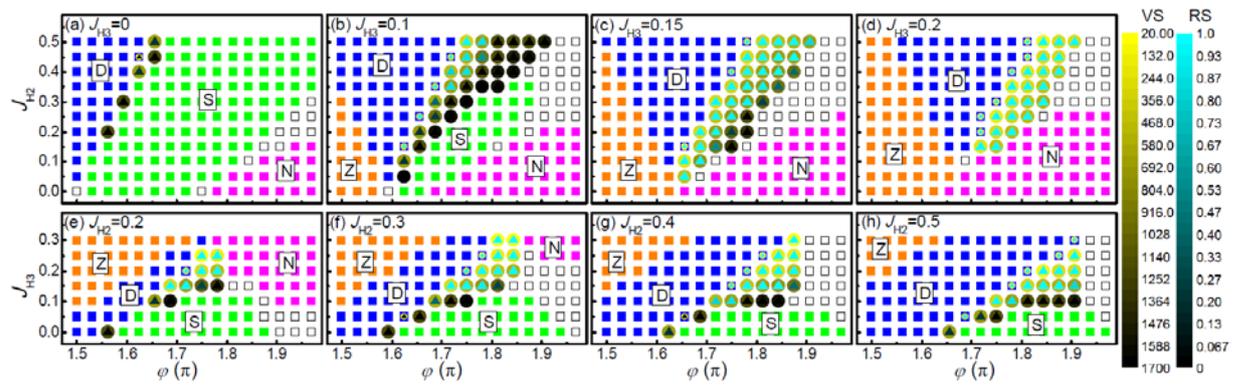

Fig. 3

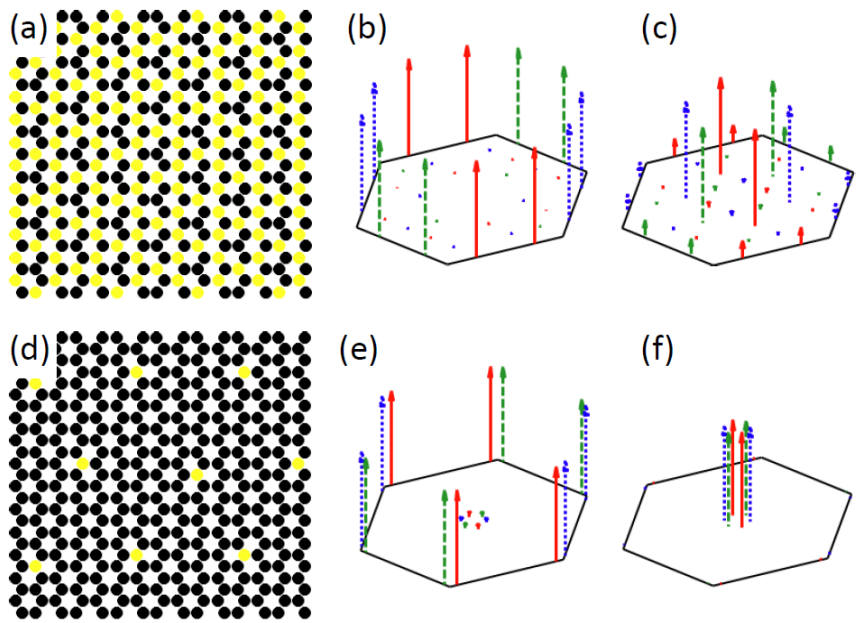

Fig. 4

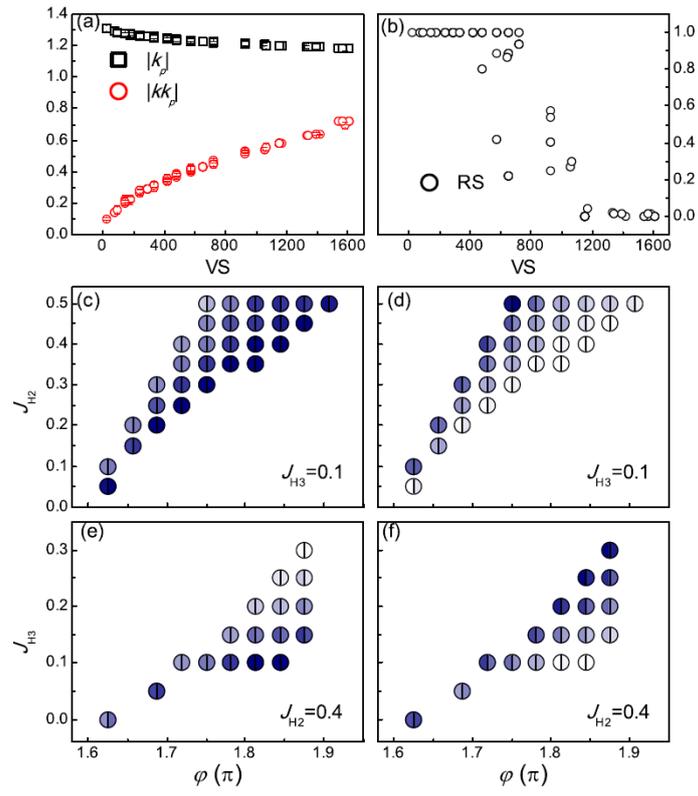

Fig. 5

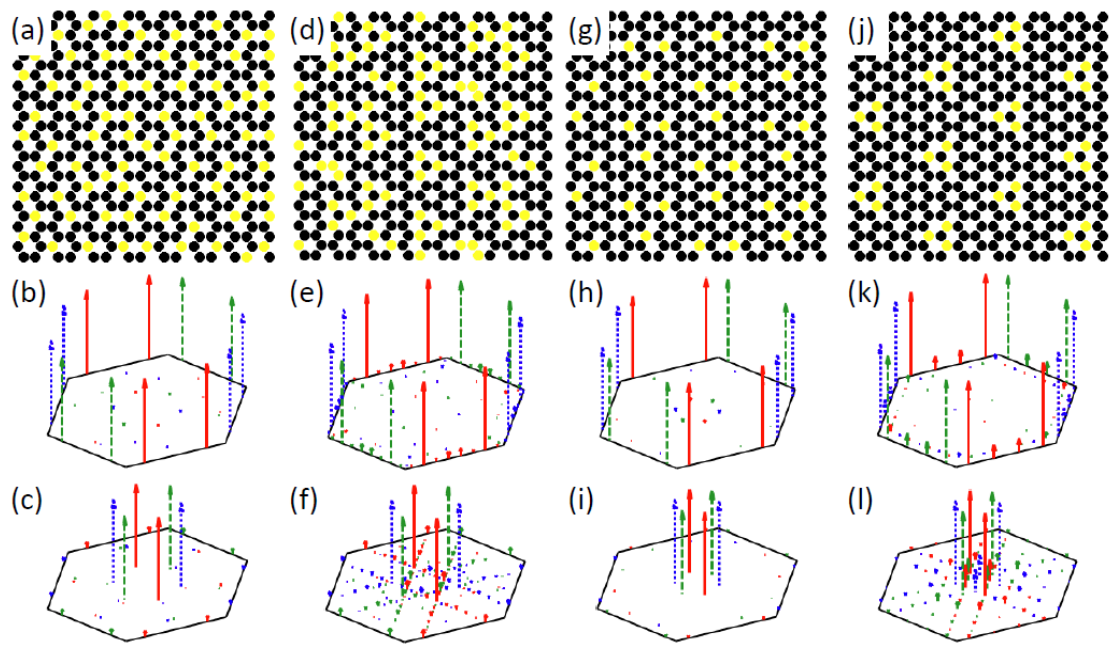

Fig. 6

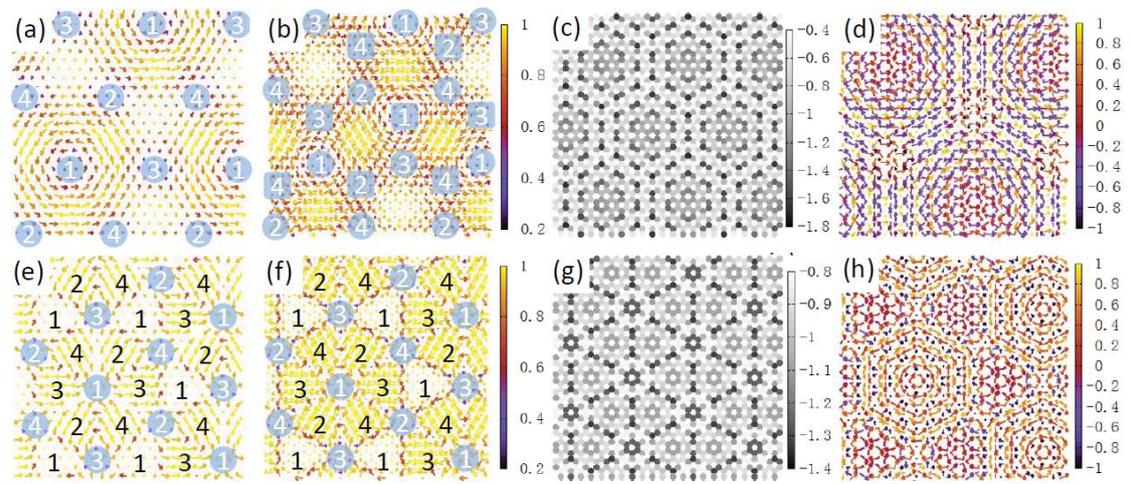